\documentclass[12pt]{article}
\usepackage{bm}
\usepackage{epsf}
\usepackage{graphicx}

\begin{document}

\begin{center}{\Large \bf Quantum evolution of Universe in the  constrained
quasi-Heisenberg picture:  from quanta to classics?}
\end{center}

\bigskip

\begin{center}

\normalsize\bf

S. L. Cherkas
\footnotemark[1]\footnotetext[1]{cherkas@inp.minsk.by},

\it Institute for Nuclear Problems, Bobruiskaya 11, Minsk 220050,
Belarus \\

\vspace{5pt}

\bf V.L. Kalashnikov\footnotemark[2]\footnotetext[2]{
v.kalashnikov@tuwien.ac.at}

\it Institut f\"{u}r Photonik, Technische Universit\"{a}t Wien,\\
Gusshausstrasse 27/387, Vienna A-1040, Austria

\end{center}

\begin{abstract}
The quasi-Heisenberg picture of minisuperspace model is
considered. The  suggested scheme consists in quantizing of the
equation of motion and interprets all observables including the
Universe scale factor as the time-dependent (quasi-Heisenbeg)
operators acting in the space of solutions of the Wheeler--DeWitt
equation. The Klein-Gordon normalization of the wave function and
corresponding to it quantization rules for the equation of motion
allow a time-evolution of the mean values of operators even under
constraint $H=0$ on the physical states of Universe. Besides, the
constraint $H=0$ appears as the relation connecting initial values
of the quasi-Heisenbeg operators at $t=0$. A stage of the
inflation is considered numerically in the framework of the
Wigner--Weyl phase-space formalism. For an inflationary model of
the ``chaotic inflation'' type it is found that a dispersion of
the Universe scale factor grows during inflation, and thus,  does
not vanish at the inflation end. It was found also, that the ``by
hand'' introduced dependence of the cosmological constant from the
scale factor in the model with a massless scalar field leads to
the decrease of dispersion of the Universe scale factor. The
measurement and interpretation problems arising in the framework
of our approach are considered, as well.
\end{abstract}

\begin{center}

{\small Key words: quantization of the equations of motion,
\\quantum stage of inflation, scale factor dispersion}
\end{center}

\newpage

\section{Introduction}

COBE  \cite{cobe}, WMAP \cite{wmap} and other experiments on
measurements of the cosmic microwave background anisotropy have
inspired a lot of the works on the classical inflationary
potential reconstruction (see, for example, Ref. \cite{rec}).
However, it is generally accepted that the quantum effects have to
be taken into account at initial stage of the cosmological
evolution. The question arises, at what stage of the Universe
evolution the classical description is applicable. A simplest
possibility to clear up this is to built the Heisenberg picture of
a minisuperspace model and to calculate the mean values and the
dispersions of observables. If at some moment of cosmological time
$t$, we would reveal that for the operators $\hat A(t)$ and $\hat
B(t)$ describing the Universe dynamics (they can be Universe scale
factor, value of scalar field etc.) the relation $<\hat A(t)\hat
B(t)>\approx <\hat A(t)><\hat B(t)>$ is satisfied with a
sufficient accuracy, then we would change the operators in the
operator equations by their mean values and, hence, consider the
Universe classically. Appearance of a classical world in quantum
cosmology is widely discussed (see Refs. \cite{hab1,hab2}, and
citation therein). As a rule, the Wigner function served as a
diagnostic tool for the problem. In quantum region the Wigner
function is highly oscillating and has no classical limit in the
general case. But under evaluation of the mean value its
oscillations are averaged. Thus, the method considering the
observable mean values and dispersions seems to be more
straightforward for analyzing of transition to the classics than a
direct analysis of the Wigner function.

The first step is choosing of an appropriate quantization scheme.
A variety of the quantization schemes for a minisuperspace model
can be roughly divided in two classes: imposing the constraints i)
``before quantization'' \cite{hen,marco} and ii) ``after
quantization'' \cite{wheel,witt} (see also reviews comparing both
approaches \cite{barv,sim}).

In the former, the constraints are used to exclude ``nonphysical''
degrees of freedom. This allows then constructing Hamiltonian
acting in the reduced ``physical'' phase space. In such models,
Universe dynamics is introduced by the  time-depended gauge. Gauge
of this type should identify the Universe scale factor with the
prescribed monotonic function of time \cite{barv2,gitm}. This
results in a non-vanishing and generically non-stationary
Hamiltonian of the system and, thus, in the equations of motion.
Such a procedure cannot be wholly satisfactory, since it requires
to introduce a priori arbitrary function and does not allow
considering the Universe scale factor as quantum observable.

The alternative schemes prefer imposing the constraint ``after
quantization". This leads to the Wheeler--DeWitt equation on
quantum states of Universe \cite{wheel,witt}. We believe that it
is most correct description of the quantum Universe. Nevertheless,
a problem of extraction of the information about the Universe
evolution in time remains, because there is no an explicit
``time'' in the corresponding Wheeler--DeWitt equation. This
inspires discussions about ``time disappearance'' and
interpretation of the wave function of Universe \cite{hall,hal1}.
Possible solutions and interpretations of this problem like to
introduce time along the quasi-classical trajectories, or
subdivide Universe into classical and quantum parts have been
offered \cite{vil}.

Our point of view is that one can solve the "problem of time"
radically, without appealing to the quasiclassics.

Let us note that i) for some  observable $A$ the commutators $[A,
H]$ are nontrivial, i.e. the equation of motion  remain in force
even in ordinary Heisenberg picture , ii) absence of evolution of
the mean values can be proved only in the Schr\"{o}dinger
normalization of the Universe wave function. Namely, for evolution
of mean value of some Heisenberg operator $\hat A$ we have
\begin{equation}
<A(t)>=<\psi|e^{i\hat Ht}\hat A e^{-i\hat Ht} |\psi>. \label{1}
\end{equation}
Let $H$ contains differential operators like $\frac{
\partial^2}{\partial a^2 }$ or  $\frac{
\partial^2}{\partial \phi^2 }$.
Assuming that the wave function $\psi(a,\phi)$ obeys $\hat
H\psi(a,\phi)=0$ (i.e. it is ``on shell'') and is normalized in
the Schr\"{o}dinger style, one can move
$\frac{\partial^2}{\partial a^2}$ to the left side by habitual
operation $<\psi|\frac{\partial^2}{\partial a^2 }=<
\frac{\partial^2}{\partial a^2 }\psi|$ through integration by
parts (and do the same for $\frac{
\partial^2}{\partial \phi^2 }$ ). As a result,
$<\psi|\hat H=<\hat H\psi|=0$ and one finds no an evolution of the
mean values with the time. iii) However, the Universe wave
function cannot be normalized in the Schr\"{o}dinger style if the
most natural Laplacian like operator ordering in the
Wheeler-DeWitt equation is chosen (for closed Universe and
unnatural operator ordering Schr\"{o}dinger's norm can be archived
\cite{haw}).

If the wave function is unbounded along one of the variables (e.g.
$a$ variable)  its normalization differs from the Schr\"{o}dinger
one and absence of evolution of the operator mean values can not
be proven. It gives a hope that some Heisenberg-like picture is
possible for the Klein--Gordon normalization. Certainly, the
ordinary Heisenberg operators are not suitable for this aim
because they are not Hermite in the normalization above.

Also, it should be  mentioned, there are the works where the
Schr\"{o}dinger normalization for the ``off shell'' states (i.e.
not obeying the constraints) has been used
\cite{kag,mil,geor,weist}.  In Ref. \cite{kag} after evaluation of
the mean values of the operators, the proceeding to limit of the
``on shell'' states (which satisfy the constraints) leads the
time-dependence of the expectation values of some operators.
Another procedure has been used in Refs. \cite{mil,geor}, where
the constraint is considered as an equation connecting expectation
values of the operators.

The paper is organized as follows:
 in section 2, origin and
description of our quantization scheme\footnote{This quantization
scheme has many common features with a model of the
relativistic-particle-clock (i.e. particle having its own clock,
for instance, radioactive particle) \cite{clock}.} (i.e.
quantization rules for the equations of motion and formula for
evaluation of the mean values) are expounded. Quantization rules
for the quasi-Heisenberg operators are defined consistently with
choice of the hyperplane used for normalization of solutions of
the Wheeler--DeWitt equation in the Klein--Gordon style.

In section \ref{two}, the approximate solution is obtained
numerically for the quasi-Heisenberg operators and  the
corresponding mean values are evaluated and discussed. Transition
to Universe having negligible dispersion of the scale factor is
discussed in section \ref{third}. In section \ref{measurem}, the
measurement and interpretation problems in the quantum Universe
are discussed.

\section{Quantization rules, operator equations of motion, mean values evaluation}

Let us start from the Einstein action for a gravity and an
one-component real scalar field:
\begin{equation}
S=\frac{1}{16\pi G}\int d^4 x\sqrt{-g}R+\int d^4 x\sqrt-g[\frac{1}{2}%
(\partial_\mu\phi)^2-V(\phi)], \label{deistv}
\end{equation}
\noindent where $R$ is the scalar curvature and $V$ is the matter
potential which includes a possible cosmological constant
effectively. We restrict our consideration to the homogeneous and
isotropic metric:
\begin{equation}
ds^2=N^2(t)dt^2-a^2(t)d\sigma^2.
\end{equation}
Here the lapse function  $N$ represents the general time
coordinate transformation  freedom. For the restricted metric the
total action becomes
\begin{eqnarray}
S=\Omega\int N(t)\biggl\{ \frac{3}{8\pi G}a\left(
{\mathcal{K}}-\frac{\dot{a}^2}{N^2(t)} \right)
+\frac{1}{2}a^3\frac{\dot{\phi}^2}{N^2(t)}- a^3V(\phi)\biggr\}dt,
\label{action}
\end{eqnarray}
\noindent where $\mathcal{K}$ is the signature of the spatial
curvature, and $\Omega$ is the constant defining volume of the
Universe. It is equal to $2\pi^2$ for the closed Universe and is
infinite for the flat and open ones. For quantization of the flat
and open Universes, $\Omega$ should be some properly fixed
constant. It is suggested that some fluctuation, from which the
Universe arises, can be approximately considered as isolated,
having no local degrees of freedom and obeying dynamics of the
uniform and isotropic Universe. Constant $\Omega$, corresponding
to the "volume" occupied by this fluctuation is to be such that
the value of $\Omega\, a^3_{today}$ is greater than the visible
part of Universe, which is known to be isotropic and uniform.
Further we set $\Omega$ to unity, i.e. approximately $1/18$ part
from the volume of the closed Universe.

The action (\ref{action}) can be obtained from the following
expression by varying on $p_a$ and $p_\phi $:
\begin{eqnarray*}
S=\int\biggl\{p_\phi\dot{\phi}+p_a\dot{a}-N(t) \biggl(-\frac{3a{\mathcal{K}}}{8\pi G}%
-\frac{8\pi G\,p_a^2}{12a}
+\frac{p_\phi^2}{2a^3}+a^3V(\phi)\biggr) \biggr\}dt.
\end{eqnarray*}
Varying on $N$ gives the constraint
\begin{eqnarray}
H=-\frac{3a{\mathcal K}}{8\pi G}-\frac{8\pi G\,p_a^2}{12a}+\frac{p_\phi^2}{2a^3}%
+a^3V(\phi)=0. \label{constr1}
\end{eqnarray}
This constraint turns into the Wheeler--DeWitt equation $ \hat
H\psi(a,\phi)=0$ after quantization: $[\hat a,\hat p_a]=-i$,
$[\hat \phi,\hat p_\phi]=i$.

Attempts to modernize or remove the constraint equation can be
justified within the framework of theories implying existence of
some preferred system of reference. For instance, the Logunov's
relativistic theory of gravity \cite{log}, which gives an adequate
description of the Universe expansion \cite{kalash}, allows
omitting the constraint \cite{lasuk}. However, here we shall keep
to the General Relativity.

Let us first consider the flat Universe (${\mathcal K }=0$) with
$V(\phi)=0$ (corresponding Hamiltonian is $H_0=\frac{p_\phi^2}{2
\,a }+\frac{p_\phi^2}{2\, a}$ in the units $4\pi G/3=1$ ).

Procedure, which is invariant under general coordinate
transformations consists in postulating the quantum Hamiltonian
\cite{witt1}:
\begin{equation}
\hat H_0=\frac{1}{2}g^{-\frac{1}{4}}\hat p_\mu g^{\frac{1}{2}}
g^{\mu\nu}\hat p_\nu g^{-\frac{1}{4}},
\end{equation}
where $\hat p_\mu=-i g^{-\frac{1}{4}}\, \frac{\partial}{\partial x^\mu}\, g^{%
\frac{1}{4}}=-i\left(\frac{\partial}{\partial x^\mu}+\frac{1}{4}(\frac{%
\partial \ln g}{\partial x^\mu})\right)$. For our choice of variables $%
x^{\mu}=\{a,\phi\}$, $p_\mu=\{-p_a,p_\phi\}$, the metric has the
form: $ g^{\mu\nu}\ = \left(
\begin{array}{cc}
-\frac{1}{a}\; & \;0 \\
\, & \, \\
\,\,0\; & \;\frac{1}{a^3}%
\end{array}
\right), $
so that $g=\det|g_{\mu\nu}|=a^4$, $\hat p_a=i\left(\frac{\partial}{\partial a }+\frac{1}{a}%
\right)$. Then the Hamiltonian is
\begin{equation}
\hat H_0=-\frac{1}{4}\left(\hat p_a^2\frac{1}{a}+\frac{1}{a}\hat
p_a^2\right)+\frac{\hat p_\phi^2}{2 a^3}= \frac{1}{2 a^2}\frac{\partial}{%
\partial a} a\frac{\partial}{\partial a}-\frac{1}{2 a^3}\frac{\partial^2}{%
\partial \phi^2}.
\end{equation}

Explicit expression for the wave function satisfying $\hat
H_0\psi=0$ is
\[
\psi_k(a,\phi)=a^{\pm i|k|}e^{ik\phi}.
\]

Exactly as in the case of the Klein--Gordon equation, we should
choose only the positive frequency solutions \cite{vil}. Thus, the
wave packet
\begin{equation}
\psi(a,\phi)= \int c(k)\frac{a^{-
i|k|}}{\sqrt{4\pi|k|}}e^{ik\phi}d k \label{pack}
\end{equation}
will be normalized by
\begin{equation}
ia\int\left(\frac{\partial \psi}{\partial a}\psi^*-\frac{\partial \psi^*}{%
\partial a}\psi \right)d\phi=\int c^*(k)c(k)dk=1,  \label{nm}
\end{equation}
where some hyperplane $a=const$ is chosen.

 Now we have to quantize the classical equations of
motion
\begin{equation}
{p}_\phi^{\bm\cdot}(t)=0, ~~~(a^3(t))^{\bm\cdot}={3} p_a a,~~~
(p_a a)^{\bm\cdot}=-3 H_0 \label{eq}
\end{equation}
obtained from the classical  hamiltonian $H_0$ by taking  Poisson
brackets $\dot A= \{H,A\}$, where
\begin{equation}
\{A,B\}=\frac{\partial A}{\partial p_\mu}\frac{\partial
B}{\partial x^\mu}- \frac{\partial A}{\partial
x^\mu}\frac{\partial B}{\partial p_\mu}=\frac{\partial A}{\partial
p_\phi}\frac{\partial B}{\partial \phi}- \frac{\partial
A}{\partial \phi}\frac{\partial B}{\partial p_\phi}-\frac{\partial
A}{\partial p_a}\frac{\partial B}{\partial a}+ \frac{\partial
A}{\partial a}\frac{\partial B}{\partial p_a}.
\end{equation}

For quantization it is sufficient to specify commutation relation
for operators at initial moment of time $t=0$. According to the Dirac quantization procedure \cite{dirac},
besides the hamiltonian constraint $\Phi_1=-\frac{%
p_a^2}{a}+\frac{p_\phi^2}{a^3}$ (see (\ref{constr1})), we have to
set some additional gauge fixing  constraint, which can be chosen
in our case as $\Phi_2=a=const$, because the hyperplane $a=const$
is chosen earlier for the normalization of the wave function in
the Klein-Gordon style. Besides the ordinary Poisson brackets the
Dirac brackets have to be introduced:
\begin{equation}
\{A,B\}_D=\{A,B\}-\{A,\Phi_i\}(C^{-1})_{ij}\{\Phi_j,B\},
\end{equation}

\noindent where $C$ is the nonsingular matrix with the elements $%
C_{ij}=\{\Phi_i,\Phi_j\}$ and $C^{-1}$ is the inverse matrix.
Quantization consists in postulating the commutator relations to
be equal to the Dirac brackets with the variables replaced   by
operators:
\begin{equation}
[\hat{\bm\eta},\hat{\bm\eta}^\prime]=-i\{\bm \eta, \bm \eta^\prime \}_D%
\biggr|_{\bm \eta\rightarrow\hat{\bm \eta}}.
\end{equation}
Here $\bm \eta$ implies set of the canonical variables $p_\mu,
x^\nu$. In contrast to the usual formalism of Refs.
\cite{hen,gitm,tyu,shab}, we postulate to impose constraints
$\Phi_1=0$ and $ \Phi_2=0$  at only hyperplane $t=0$. Consequently
the quasi-Heisenberg operators obey the commutation relations
obtained from the Dirac quantization procedure at the initial
moment $t=0$. Direct evaluation gives
\begin{eqnarray}
[{\hat p}_a(0),{\hat a}(0)]=0,~~~
{[\hat p_\phi(0),\hat a(0)]=0},  \nonumber \\
{[{\hat p}_\phi(0),{\hat \phi}(0)]=-i},~~~
{[{\hat p}_a(0),\hat \phi(0)]=-i\frac{{\hat p}_\phi(0)}{{\hat p}_a(0){\hat a}%
^2(0) }}.  \label{rec0}
\end{eqnarray}

One has to solve Eqs. (\ref{eq}) with the given initial
commutation relations. In contract to the ordinary  Heisenberg
operators, the quasi-Heisenberg operators do not conserve their
commutation relations during evolution. The commutation relations
(\ref{rec0}) can be satisfied through
\begin{equation}
\hat a(0)=const=a, ~~~ \hat p_\phi(0)=\hat p_\phi, ~~~ \hat
p_a(0)=|\hat p_\phi|/a , ~~~ \hat \phi(0)=\phi,  \label{rec1}
\end{equation}
where $\hat p_\phi=-i\frac{\partial}{\partial \phi}$. Variable $a=\hat a(0)$ is $c$%
-number now because it commutes with all operators \cite{shab}.
Solutions of Eqs. (\ref{eq}) are
\begin{eqnarray}
\hat p_\phi(t)=\hat p_\phi,  ~~~ \hat a^3(t)=a^3+3|\hat
p_\phi|t,~~~ \hat p_a(t)=\frac{|\hat p_\phi|}{(a^3+3|\hat
p_\phi|t)^{1/3}},\nonumber \\ \hat \phi(t)=\phi+\frac{\hat
p_\phi}{3|\hat p_\phi|}\ln(a^3+3|\hat p_\phi|t)-\frac{\hat
p_\phi}{|\hat p_\phi|}\ln\,a. \label{oper}
\end{eqnarray}

We imply that these quasi-Heisenberg operators act in the Hilbert
space with the Klein-Gordon scalar product. Expression for the
mean value of an observable is
\begin{eqnarray} <\hat A(t)>=i\,a\int \biggl( \psi^*(a,\phi)
D^\frac{1}{4} \hat A(t)D^{-\frac{1}{4}} \frac{\partial}{\partial
a} \psi(a,\phi)~~\nonumber
\\
~~~~~~~~~~~~~~~~- \frac{\partial}{\partial a
}\psi^*(a,\phi)D^{-\frac{1}{4}}\hat
A(t)D^{\frac{1}{4}}\psi(a,\phi)\biggr)d\phi \biggr| _{a\rightarrow
0},~~ \label{nmm}
\end{eqnarray}
where operator $D=-\frac{\partial^2}{\partial
\phi^2}+2\,a^6V(\phi)$ (since $a\rightarrow 0$ the $V$-term can be
omitted in the expression for $D$). Eq. (\ref{nmm}) is particular
case of that suggested in Ref. \cite{ali}, where an one-particle
picture of the Klein-Gordon equation in the Foldy-Wouthausen
representation has been considered. The adequacy of this
definition  can been seen in the momentum
representation of the $\phi$ variable, where $\hat p_\phi=k$ and $%
\hat \phi= i\frac{\partial}{\partial k}$. Then Eq. (\ref{nmm})
gives
\begin{eqnarray}
 <\hat A(t)>=\int {a}^{i|k|}c^*(k)\hat A(t,\hat \phi,k,a){a}^{-i|k|}c(k)dk\biggr|_{a\rightarrow
0}, \label{defali}
\end{eqnarray}
which is similar to the ordinary quantum mechanical definition and
certainly possesses hermicity.

 Evaluation of the mean value $\hat a^3(t)$ given by (\ref{oper}), (\ref{defali}) over the wave packet
 (\ref{pack}) reads
\begin{equation}
<a^3(t)>=3t\int|k||c(k)|^2 dk.
\end{equation}

Next  quantity is the mean value of the scalar field
$<\hat\phi(t)>$:

\begin{eqnarray}
   <\hat\phi(t)>_{a}=\int {a}^{i|k|}c^*(k)\left(
i\frac{\partial}{\partial k}+\frac{k}{3|k|}\ln\left( a^3+3|k|t
\right)-\frac{k}{|k|}\ln a \right)a^{-i|k|}c(k) dk
\nonumber\\
=\int\left( c^*(k)i\frac{\partial}{\partial
k}c(k)+\frac{k}{3|k|}\ln\left(a^3+3|k|t\right)
|c(k)|^2\right)dk.~~ \label{canc}
\end{eqnarray}

Brackets $<\dots>_a$ with the index $a$ in (\ref{canc}) mean that
$a$ is not equal to zero yet (compare with Eq. (\ref{defali})). A
remarkable property of Eq. (\ref{canc}) is that the term $-\frac{k}{|k|%
}\ln a$ cancels the term arising from the differentiation: $a^{i|k|}i\frac{%
\partial}{\partial k}a^{-i|k|}= \frac{k}{|k|}\ln a$. Thus, after $%
a\rightarrow 0 $ one can obtain
\begin{equation}
<\hat \phi(t) >=\int\left(\frac{k}{3|k|}\ln(3|k|t)|c(k)|^2 +
c^*(k)i \frac{\partial}{\partial k}c(k)\right)dk.  \label{phi_1}
\end{equation}
Cancellation of the terms divergent under $a\rightarrow0$ in the
mean values of the quasi-Heisenberg operators is a general feature
of the theory. As a result, it is possible to evaluate, for
instance,
\begin{equation}
<\hat \phi^2(t)
>=\int\biggl(\frac{1}{9}\ln^2(3|k|t)|c(k)|^2-c^*(k)\frac{\partial^2}{\partial
k^2}c(k)\biggr)dk. \label{phi2}
\end{equation}

One should not confuse the divergence at $a\rightarrow 0$ arising
under evaluation of the mean values with the singularity at
$t\rightarrow 0$. The mean values of operators, which are singular
at $t\rightarrow0 $ in the classical theory remain singular also
in the quantum case. According to (\ref{oper}), (\ref{nmm}) the
way to avoid a singularity is to guess, that the Universe
evolution began from some ``seed'' scale factor $a_0$. Then in the
expression for a mean value, one has to assume $a\rightarrow a_0$
instead of $a\rightarrow 0$. But, the mathematics is simplified
greatly namely at $a\rightarrow 0$, because the use of
asymptotical value of the wave function is possible in this case.

One more kind of the infinity can be found in  Eqs. (\ref%
{phi_1}), (\ref{phi2}): for $c(k)$, which does not tend to zero at
small $k$, the mean values of $\phi(t)$ and $\phi^2(t)$ diverge.
This is a manifestation of the well-known infrared divergency of
scalar field minimally coupled with gravity. Thus, not all
possible $c(k)$ are suitable for construction of the wave packets.

Let us consider Hamiltonian, containing the cosmological constant
$V_0$:
\begin{equation}
H=H_0+a^3V_0.  \label{ham_v}
\end{equation}
Explicit solution for the wave function $\hat H\psi=0$ has the
form
\begin{equation}
\psi_k(a,\phi)=\left(\frac{18}{V_0}\right)^{\frac{i|k|}{6}} \Gamma(1-\frac{i|k|%
}{3})J_{-\frac{i|k|}{3}}(\frac{\sqrt{2V_0}}{3}a^3)e^{i k\phi},
\label{ww}
\end{equation}
where $\Gamma(z)$ is the Gamma function and $J_\mu(z)$ is the
Bessel function. The wave function (\ref{ww}) tends to
$a^{-i|k|}e^{i k\phi}(1+O(a^6))$ asymptotically under
$a\rightarrow0$. Then for evaluation of the mean values according
to (\ref{defali}), we can always build the wave packet
$a^{-i|k|}e^{i k \phi }$ from solutions of the free
Wheeler--DeWitt equation and do not encounter with a problem of
negative frequency solutions.  The argumentation holds for any
potential $V(\phi)$, because it contributes into the Hamiltonian
as a term multiplied by $a^3$.

Equations of motion obtained from the classical Hamiltonian
(\ref{ham_v}) are
\begin{equation}
(a^3(t))^{\bm\cdot}={3}\hat p_a a,~~~ (p_a a)^{\bm\cdot}=3V_0a^3-3
H_0, \label{a2} ~~~ (V_0a^3-H_0)^{\bm\cdot}=6 \,V_0\, p_a a.
\label{a3}
\end{equation}
The additional term $a^3V_0$ does not change relations
(\ref{rec0}) required for the quantization procedure. Only
expression for $\hat p_a(0)$ changes in (\ref{rec1}): $\hat
p_a(0)=\sqrt{\frac{\hat p_\phi^2}{a^2}+2V_0a^4}$.

Finally we arrive to
\begin{equation}
\hat a^3(t)=a^3+3|\hat p_\phi|\frac{\sinh(t\sqrt{18 V_0})}{\sqrt{18 V_0%
}} + a^3(\cosh(t\sqrt{18 V_0})-1).  \label{os}
\end{equation}

\noindent Evaluation of the mean values according to Eq.
(\ref{nmm}) leads to
\begin{eqnarray*}
<\hat a^3(t)>=\frac{\sinh (3\,\sqrt{2V_0}\,t )} {\sqrt{2V_0}}\int
|k||c(k)|^2dk, \\
<\hat a^6(t)>=\frac{\left({\sinh (3\,\sqrt{2V_0}\,t )}\right)^2}{2\,V_0%
}\int k^2 |c(k)|^2dk.
\end{eqnarray*}
One can see, that the dispersion $\sqrt{<\hat a^6>-<\hat
a^3>^2}/<\hat a^3>$ does not depend on $t$. Thus, the evolution of
Universe remains quantum during all time in the model with a
cosmological constant.

This results from an absence of some scale length in the model
with cosmological constant (besides the natural Plank length).
Such a length can appear due to some mechanism reducing a
cosmological constant during the Universe evolution. In the next
section one of the possible mechanisms, namely an inflation
derived by the quadratic potential of the scalar field, is
considered.

\section{Operator equations for the quadratic inflationary potential and
Wigner-Weyl evolution of the minisuperspace} \label{mini}
\label{two}

As it has been discussed, the quantization procedure consists in
quantization of the equations of motion, i.e. considering  them as
the operator equations. These equations have to be solved with the
initial conditions obeying to the constraint at $t=0$. For the
Hamiltonian
\begin{equation}
H=-\frac{p_a^2}{2 a}+\frac{p_\phi^2}{2 a^3}+a^3\frac{ m^2
\phi^2}{2}
\end{equation}
we have the equations:
\begin{eqnarray}
\ddot a=-\frac{3}{2}a\dot \phi^2-\frac{\dot a^2}{2
a}+\frac{3}{2}a\, m^2 \phi^2, \nonumber \\\ddot \phi=-3\frac{\dot
a}{a}\dot \phi-m^2 \phi \label{infl1}
\end{eqnarray}
and the constraint:
\begin{equation}
-{\dot a^2}{a}+{\dot \phi^2}{a^3}+a^3 m^2\phi^2=0. \label{conop}
\end{equation}

\noindent The point means the differentiation over $t$. After
quantization, Eqs. (\ref{infl1}) lead to the equations for the
quasi-Heisenberg operators, which have to be solved with the
operator initial conditions:
\begin{eqnarray}
\hat a(0)\equiv a,~~~
 \hat \phi(0)\equiv\phi,~~~
\hat p_\phi(0)\equiv -i\frac{\partial}{\partial \phi},\nonumber\\
\dot {\hat \phi}(0)=\frac{\hat p_\phi(0)}{\hat
a^3(0)}=\frac{1}{a^3}\left(-i\frac{\partial}{\partial \phi}\right), \nonumber\\
\dot {\hat a}(0)=\hat a(0)\sqrt{{\dot {\hat \phi}^2(0) }+m^2 \hat
\phi^2(0)} =\sqrt{\frac{1}{a^4}\left(-i\frac{\partial}{\partial
\phi }\right)^2+m^2 a^2\phi^2}.
\end{eqnarray}
According to our ideology, the operator constraint (\ref{conop})
is satisfied only at $t=0$. The ordinary problem of the operator
ordering arises, because the quasi-Heisenberg operators are
noncommutative in the general case. The problem seems more
transparent if we change the variable $\hat \alpha= \ln \hat a$:
\begin{eqnarray}
\ddot{\hat\alpha} +\frac{3}{2}\dot{\hat\alpha}^{2}-\frac{3}{2}
m^2\,\hat\phi^2+\frac{3}{2}{\dot{\hat\phi}}^{2}=0,\nonumber\\
{\ddot{\hat\phi}} +\frac{3}{2}\left(\dot{\hat
\alpha}\dot{\hat\phi}+\dot{\hat\phi}\dot{\hat\alpha}\right)+m^2\hat\phi=0,
\label{syst}
\end{eqnarray}
 where the symmetric ordering is used.
The system (\ref{syst}) has to be solved with the initial
conditions: $\hat \phi(0)=\phi$, $\hat \alpha(0)=\ln a$, $\hat
{\dot \phi}(0)=\frac{1}{a^3}\left(-i\frac{\partial}{\partial
\phi}\right)$, $ \dot{\hat
\alpha}(0)=\sqrt{m^2\phi^2+\frac{1}{a^6}\left(-i\frac{\partial}{\partial
\phi }\right)^2} $.

\begin{figure}[h]
\begin{center}
\epsfbox{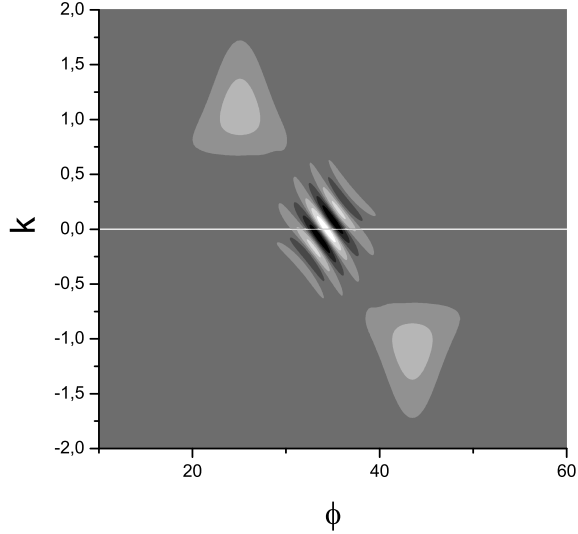}
\end{center}
\caption{\label{fig1} Contour-plot of the Wigner function of
Universe for $c(k)=e^2\left(\frac{2}{\pi}\right)^{1/4}\exp(ik
\phi_0- k^2-1/k^2)$ at $a=10^{-4}$.}
\end{figure}

\begin{figure}[h]
\begin{center}
\epsfbox{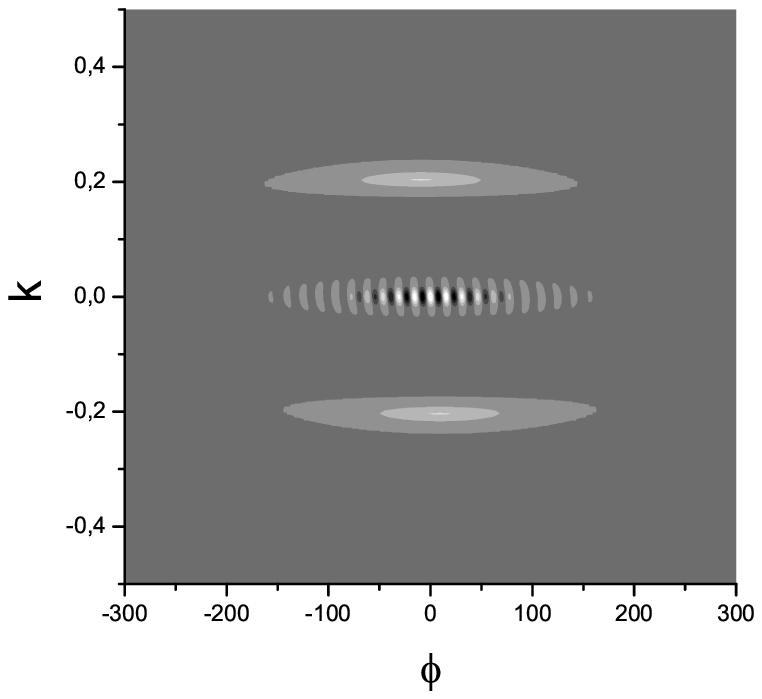}
\end{center}
\caption{\label{fig2} Contour-plot of the Wigner function of
Universe for
$c(k)=2\sqrt{5}\,e^{20\sqrt{6}}\left(\frac{3}{\pi}\right)^{1/4}\exp(-600
k^2-1/k^2)$ at $a=10^{-4}$.}
\end{figure}

The operator equations under consideration can be solved within
the framework of the perturbation theory in the first order on
interaction constant (i.e. on $m^2$). The solution in analytical
form can be found in \cite{clock}. The analytic solution is
important because it allows ensuring that the divergent terms at
$a\rightarrow 0$ cancel each other under calculations of the mean
values based on (\ref{defali}).

However, the most interesting is to consider an inflation at its
late stages. This requires a numerical consideration of the
operator equations and can be realized within the framework of the
Weyl-Wigner phase-space formalism \cite{groot}. Let us remind that
in this formalism every operator acting on $\phi$ variable has the
Weyl symbol: ${\mathcal W }[\hat A]=A(k,\phi)$. For instance, the
simplest Weyl symbols in our case are: $\mathcal
W[-i\frac{\partial }{\partial \phi}]=k$, ${\mathcal
W[\phi]=\phi}$. Weyl symbol of the symmetrized product of
operators reads
\begin{eqnarray}
   W[\frac{1}{2}(\hat A\hat B+\hat B\hat
A)]=\cos\biggl(\frac{\hbar}{2}\frac{\partial }{\partial
\phi_1}\frac{\partial }{\partial k_2}
-\frac{\hbar}{2}\frac{\partial }{\partial \phi_2}\frac{\partial
}{\partial k_1}\biggr)A(k_1,\phi_1)B(k_2,\phi_2)\biggr|_{\begin
{array}{c}
 ^{k_1=k_2=k} \\
^{\phi_1=\phi_2=\phi}
\end {array}},~
\label{cos}
\end{eqnarray}

 where the Planck constant is restored only to point the
order of cosine expansion.

Let us consider the Weyl transformation of Eqs. (\ref{syst}) and
expand the Weyl symmetrized product of operators up to
second-order in $\hbar$. This results in:

\begin{eqnarray}
   \partial^2_t
\alpha+\frac{3}{2}\biggl((\partial_t\alpha)^2
+\frac{\hbar^2}{4}(\partial_{k}\partial_{\phi}\partial_t\alpha)^2
-
\frac{\hbar^2}{4}(\partial^2_{\phi}\partial_t\alpha)(\partial^2_{k}\partial_t\alpha)
\biggr)+\frac{3}{2}\biggl((\partial_t\varphi)^2
+\frac{\hbar^2}{4}(\partial_{k}\partial_{\phi}\partial_t\varphi)^2
\nonumber\\
-
\frac{\hbar^2}{4}(\partial^2_{\phi}\partial_t\varphi)(\partial^2_{k}\partial_t\varphi)\biggr)-
\frac{3}{2}m^2\biggl(\varphi^2
+\frac{\hbar^2}{4}(\partial_{k}\partial_{\phi}\varphi)^2 -
\frac{\hbar^2}{4}(\partial^2_{\phi}\varphi)(\partial^2_{k}\varphi)
\biggr)=0,\nonumber\\
  \partial^2_t \varphi+3\biggl(
\partial_t\alpha\partial_t\varphi+ \frac{\hbar^2}{4}
(\partial_k\partial_\phi\partial_t\alpha)(\partial_{k}\partial_\phi\partial_t\varphi)
-\frac{\hbar^2}{8}(\partial^2_{k}\partial_t\alpha)
(\partial^2_{\phi}\partial_t\varphi)
\nonumber\\-\frac{\hbar^2}{8}(\partial^2_{\phi}\partial_t\alpha)(\partial^2_{k}\partial_t\varphi)\biggr)+
m^2\varphi=0,
\nonumber\\
\label{num}
 \end{eqnarray}

\noindent where $\alpha(k,\phi,t)$ and $\varphi(k,\phi,t)$ are the
Weyl symbols of the operators $\hat \alpha(t)$ and $\hat \phi(t)$,
respectively. These equations have to be solved with the initial
conditions at $t=0$:
\begin{eqnarray} \nonumber
  \alpha \left(k,\phi, 0 \right) = \ln a ,~~~
  \partial _t  \alpha \left(k,\phi, 0 \right)
   ={\mathcal W}\left[ \sqrt{ -{\frac{{1 }}
{{a^6 }}\frac{\partial^2}{\partial \phi^2} + m^2\phi ^2 }}\right] , \hfill \nonumber\\
  \varphi \left(k,\phi, 0 \right) = \phi ,~~\partial _t  \varphi \left(k,\phi, 0 \right) = \frac{k}
{{a^3 }}. \hfill \label{inc}
\end{eqnarray}

\noindent Weyl symbol of the square root can be expressed as
\cite{root}:
\begin{eqnarray}
{\mathcal W}\left[ \sqrt{ -{\frac{{1 }} {{a^6
}}\frac{\partial^2}{\partial \phi^2} + m^2\phi ^2
}}\right]=\frac{m^{1/2}}{\pi^{1/2}a^{3/2}}\int_0^\infty
t^{-1/2}\exp\left(-\frac{m^2\,a^6\phi^2+k^2}{m\,a^3}\tanh(t)\right)
\nonumber\\
\times \mbox{sech}(t)\left(\frac{m^2\,a^6\phi^2+k^2}{m\,a^3}
\mbox{sech}(t)^2+\tanh(t)\right)dt.~~
\end{eqnarray}
Since the mean values result from $a\rightarrow 0$, it is possible
to take simply $\partial _t \alpha \left(k,\phi, 0 \right)
   =\frac{|k|}{a^3}$.

\begin{figure*}
\includegraphics[width=10 cm]{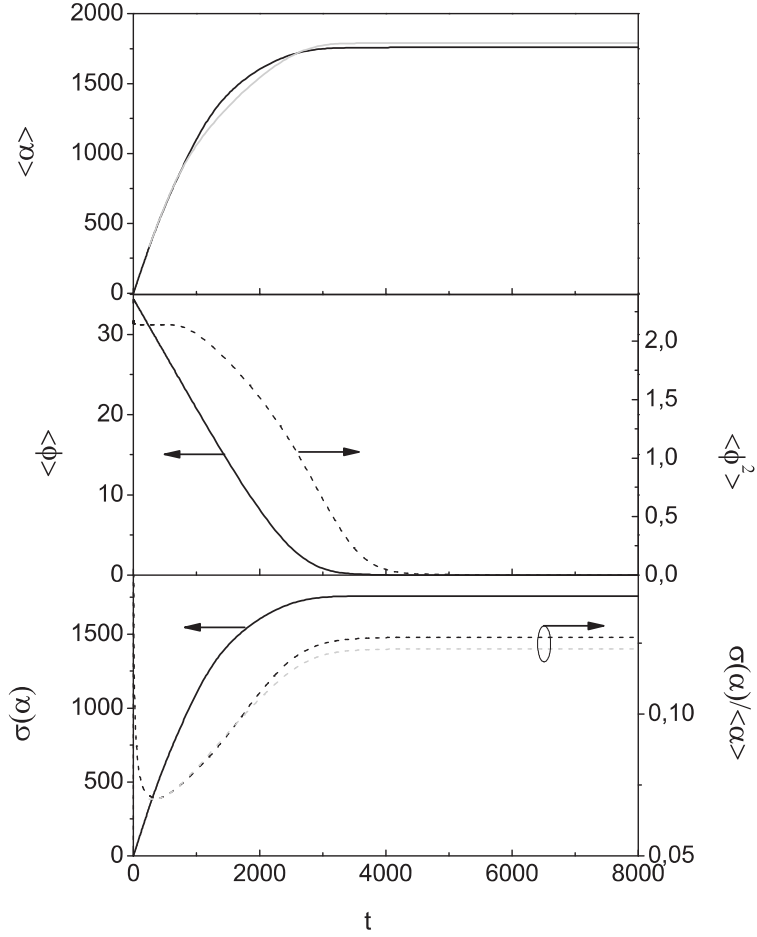} \caption{\label{fig3} Evolution of $
\left\langle \alpha  \right\rangle ,\,\left\langle {\varphi  }
\right\rangle$, dispersion $\sigma \left( \alpha  \right) = \sqrt
{\left\langle {\alpha ^2 } \right\rangle  - \left\langle \alpha
\right\rangle ^2 }$ (solid curves); $\left\langle {\varphi ^2 }
\right\rangle$ and relative dispersion $\sigma \left( \alpha
\right)/\left\langle \alpha  \right\rangle$ (dashed curves) for
$c(k)=e^2\left(\frac{2}{\pi}\right)^{1/4}\exp(ik \phi_0-
k^2-1/k^2)$. $\hbar = 0$ (black curves), $\hbar = 1$ (gray
curves).}
\end{figure*}

\begin{figure*}
\includegraphics[width=10 cm]{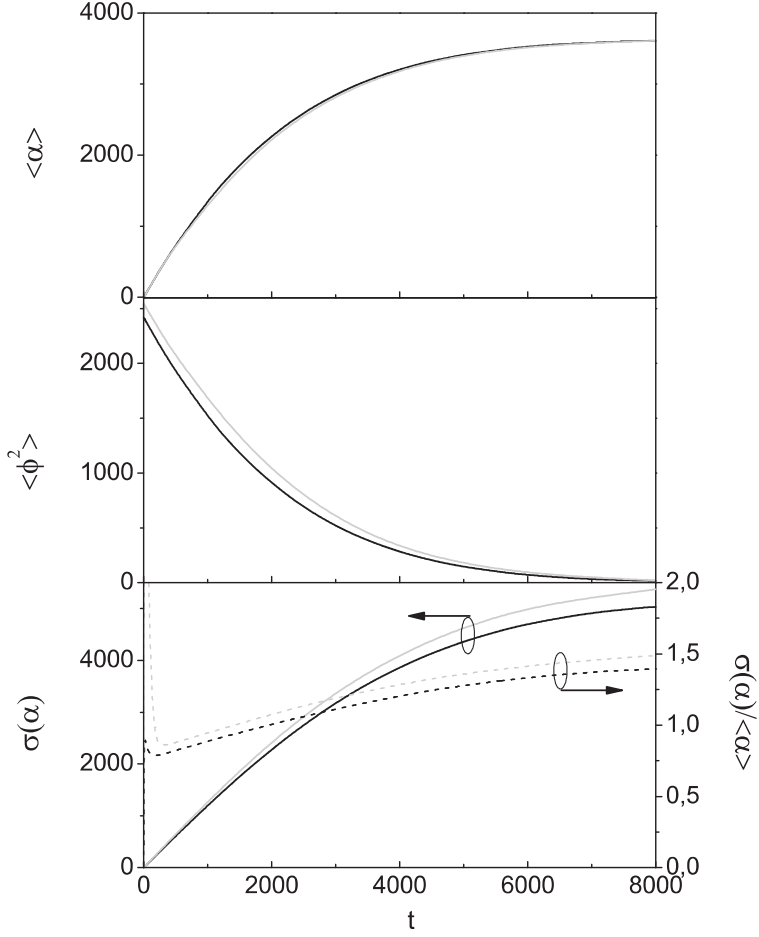} \caption{\label{fig4} Evolution of $
\left\langle \alpha  \right\rangle ,\,\left\langle {\varphi ^2 }
\right\rangle$, dispersion $\sigma \left( \alpha  \right) = \sqrt
{\left\langle {\alpha ^2 } \right\rangle  - \left\langle \alpha
\right\rangle ^2 }$ (solid curves) and relative dispersion $\sigma
\left( \alpha \right)/\left\langle \alpha  \right\rangle$ (dashed
curve) for
$c(k)=2\sqrt{5}\,e^{20\sqrt{6}}\left(\frac{3}{\pi}\right)^{1/4}\exp(-600
k^2-1/k^2)$. $\hbar = 0$ (black curves), $\hbar = 1$ (gray
curves).}
\end{figure*}

  State of Universe  is described by the Wigner function $\wp(k, \phi)$, which
is constructed on the basis of definition (\ref{nmm}) and given by

\begin{eqnarray}\label{wigner} \nonumber
   \wp \left( {k,\phi } \right) = ia\int {\left[ {\left| { - \frac{{\partial ^2 }}
{{\partial \phi ^2 }}} \right|^{ - \frac{1} {4}} \psi ^* \left(
{\phi  + \frac{u} {2}} \right)} \right]\left[ {\left| { -
\frac{{\partial ^2 }} {{\partial \phi ^2 }}} \right|^{ - \frac{1}
{4}} \frac{{\partial \psi \left( {\phi  - \frac{u} {2}} \right)}}
{{\partial a}}} \right]e^{iku} du}  -  \hfill \\
  ia\int {\left[ {\left| { - \frac{{\partial ^2 }}
{{\partial \phi ^2 }}} \right|^{ - \frac{1} {4}} \frac{{\partial
\psi ^* \left( {\phi  + \frac{u} {2}} \right)}} {{\partial a}}}
\right]\left[ {\left| { - \frac{{\partial ^2 }} {{\partial \phi ^2
}}} \right|^{ - \frac{1} {4}} \psi \left( {\phi  - \frac{u} {2}}
\right)} \right]e^{iku} du.~}  \hfill
\end{eqnarray}

\noindent or in the momentum representation of the wave function
corresponding to Eq. (\ref{pack}):
\begin{equation}
\wp(k,\phi)=\frac{1}{\pi}\int
c^*(2k-q)c(q)a^{-i|q|+i|2k-q|}e^{2i(q-k)\phi}dq.
\end{equation}

As a result of $a\rightarrow 0$, both Weyl symbols and Wigner
function diverge. In particular, when $a\rightarrow 0$ the Wigner
function becomes strongly oscillating. However, the divergences
cancel each other in
  the  expectation values, which can be
constructed in an ordinary way. For instance, expectation values
of $\alpha$ and its square are: \begin{eqnarray*} \left\langle
\alpha(t) \right\rangle = \int {dkd\phi \,\alpha(k,\phi,t)\wp
\left( {k,\phi } \right)}|_{a\rightarrow 0}\,,\\
 \left\langle
\alpha^2(t) \right\rangle = \int dkd\phi\bigl( \alpha^2
+\frac{\hbar^2}{4}(\partial_{k}\partial_{\phi}\alpha)^2 -
\frac{\hbar^2}{4}(\partial^2_{\phi}\alpha)(\partial^2_{k}\alpha)
\bigr)\wp ( {k,\phi } )\biggr|_{a\rightarrow 0}.
\end{eqnarray*}

Let us discuss the parameters of inflationary model. Quadratic
potential corresponds to the  Linde's ``chaotic inflation''
\cite{linde}. This model supposes that the value of potential at
an initial stage of the inflation has to be an order of the Planck
mass $M_p$ in fourth degree
($M_p=G^{-1/2}=\sqrt{\frac{4\pi}{3}}$). Hence, the corresponding
value of the scalar field is $\phi_0=\frac{\sqrt{2}M_p^2}{m}$.
Constant $m^2$ dictated by the COBE data is $m^2\sim
10^{-12}M_p^2=10^{-12}\frac{3}{4\pi}$. Still for the purposes of
visuality of the numerical calculations we take $m^2=1.7\times
10^{-3}$. This reflects the fact that $m^2<<M_p^2$ and the initial
scalar field is sufficiently large to provide $V^{1/4}\sim M_p$.
There are two possibility to create large scalar field: the first
one is a wave packet with the non-zero mean field $\phi_0$, for
instance, $c(k)=e^2\left(\frac{2}{\pi}\right)^{1/4}\exp(ik \phi_0-
k^2-1/k^2)$ (the corresponding Winger function is shown in Fig.
\ref{fig1}). The second one is a ``squeezed'' packet having small
uncertainty of $k$, but large square of the scalar field: for
instance,
$c(k)=2\sqrt{5}\,e^{20\sqrt{6}}\left(\frac{3}{\pi}\right)^{1/4}\exp(-600
k^2-1/k^2)$ (the corresponding Winger function is shown in Fig.
\ref{fig2}). Note that for ordinary systems, the decoherence
principle forbids the highly ``squeezed'' packets because they
should be ``collapsed'' due to interacting with environment. The
minisuperspace model takes up a little number degrees of freedom
and the other ones may serve as an  environment. So, we cannot be
fully sure that the ``squeezed'' packet is permitted.

For the both packets, the function $c(k)$ contains a multiplier
$\exp(-1/k^2)$ suppressing the infrared divergence. As a result of
the numerical solution of Eqs. (\ref{num}), the evolution of the
operators expectation values and their dispersions have be
obtained. Results are shown in Figs. \ref{fig3},\ref{fig4}.

We consider two different cases for Eqs. (\ref{num}): i) $\hbar =
0$  and ii) $\hbar = 1$ (that gives corrections to the equations
of motion taking into account the operator noncommutativity).

The main conclusion is that the dispersion of logarithm of the
Universe scale factor does not vanish during inflation. Even for
the wave packet having a large mean value of the scalar field
$\phi_0$ and a small dispersion, one can see (Fig. \ref{fig3}) the
increasing dispersion of the scale factor logarithm during
inflation without dispersion decay after the inflation end. In our
particular case (small mass of the scalar field), corrections to
the equations of motion due to noncommutativity of the
quasi-Heisenberg operators do not change the picture
qualitatively. Smallness of the corrections indicates that the
contribution of the next terms in the cosine expansion (\ref{cos})
is negligible. Thus the accurate solutions of the operator
equations of motion (\ref{syst}) for the particular set of
parameters are obtained.

\section{Decrease of the Universe scale factor dispersion
due to inflation dynamics.} \label{third}

\begin{figure}[h]
\begin{center}
\epsfxsize =10. cm \epsfbox[50 0 410 350]{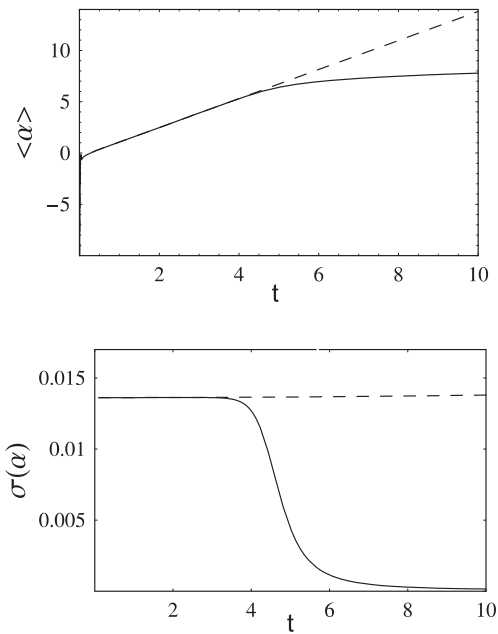}
\end{center}
\caption{\label{figx} Mean value of logarithm of the scale factor
and its dispersion for the model with the cosmological constant,
$V_0=1$, $\beta=0$ (dashed curve), and for the model
(\ref{syst111}) with the decreasing cosmological constant,
introduced ``by hands'', $V_0=1$, $\beta=10^{-8}$.}
\end{figure}

\begin{figure}[h]
\begin{center}
\epsfxsize =10. cm \epsfbox[50 0 410 350]{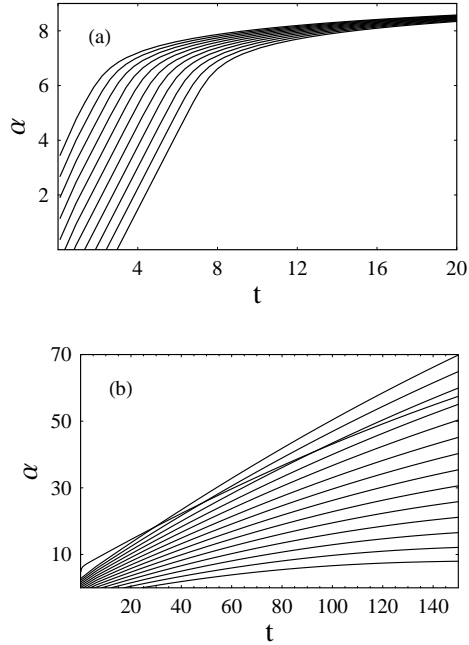}
\end{center}
\caption{\label{figxx} Classical trajectories: (a) for the model
with the decreasing cosmological constant: $V_0=1$,
$\beta=10^{-8}$, and (b) for the model with the quadratic
potential of the scalar field. Initial value of $\phi$ is fixed,
initial value of $k$ is varying.}
\end{figure}

 As we seen in the previous section, the model with one
scalar field provides Universe with growing dispersion of the
scale factor. Nevertheless, how can Universe become classical? A
possible answer is that this occurs due to decoherence. That is at
some stage of the Universe evolution, it interacts with the
environment and such interaction suppresses the quantum properties
of the system. As an ``environment'', one can consider the
remaining degrees of freedom (including matter \cite{kiefer}),
which are not taken into account in the minisuperspace model.
However, there exists one more possibility of the transition to
the classics occurring only due to the Universe dynamics without
appealing to the decoherence.

Let us consider the model with the massless scalar field but with
the introduced ``by hand'' decrease of the cosmological constant.
Hamiltonian of the model has the form:
\begin{equation}
H=-\frac{p_a^2}{2 a}+\frac{p_\phi^2}{2
a^3}+V_0\frac{a^3}{1+\beta\,a^3}, \label{infl2}
\end{equation}
where $\beta$ is some constant. The Hamiltonian suggests some
modification of the theory with cosmological constant in a sense
that the cosmological ``constant'' $V_0/(1+\beta a^3 )$ is not
zero at the small scale factors and decreases as $a^{-3}$. We
shall not discuss here what fundamental model is able to produce
such a modification. Let us only remind that the first model of
the inflation by Starobinsky  \cite{star} did not use a scalar
field.

 The corresponding equations of motion are
\begin{eqnarray}
\ddot{\alpha}
+\frac{3}{2}\dot{\alpha}^{2}+\frac{3}{2}{\dot{\phi}}^{2}-V_0\frac{3}{(1+\beta\,
e^{3\alpha})^2}=0,\nonumber\\
{\ddot{\phi}} +{3}\dot{ \alpha}\dot{\phi}=0. \label{syst111}
\end{eqnarray}
Initial conditions correspond to Eq. (\ref{inc}) apart from
\[
 \partial _t  \alpha \left(0 \right)
   = \sqrt{ -{\frac{{1 }}
{{a^6 }}\frac{\partial^2}{\partial \phi^2} + \frac{2\,V_0}{1+\beta
a^3} }}.
\]
The latter does not differ from $ \partial _t \alpha (0)
   = \frac{1}{a^3}\sqrt{-\frac{\partial^2}{\partial \phi^2}}$
   due to limit $a\rightarrow 0$ under evaluation of the mean values.

 Results of calculation are shown in Fig. \ref{figx}. One can
see that Universe becomes classical after the inflation end. A
sufficiently quick decrease of the cosmological constant allows
suppressing the dispersion of the scale factor logarithm.

The trajectories of $\alpha(k,\phi,t)$ at the fixed $\phi$ but for
the different $k$ are shown in Fig. \ref{figxx}. The trajectories
are divergent for the inflationary model with quadratic potential,
but are convergent for Eq. (\ref{infl2}). The last illustrates the
transition to classics.

We did not investigate models with the multiple scalar fields, but
considered different shapes of the potential both for the ``small
field'' and ``large field'' inflation \cite{infl}. In both cases
we had the divergent trajectories and the time-growing dispersion
of the scale factor logarithm.

\section{Measurement issue and interpretation}
\label{measurem}

\begin{figure}[h]
\hspace {2.5 cm} \epsfxsize =5. cm \epsfbox[50 100 410
450]{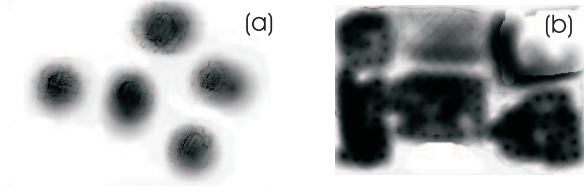} \vspace{-2.5cm} \caption{ Schematic picture: (a)
Universe after measurements by the local  observers. Scale factor
is projected to defined value in the local regions. (b) A stage,
when Universe becomes nonuniform, can be considered as a
``self-measurement''.}
 \vspace{-0.5cm}
\label{stoks}
\end{figure}

 It is often stated that the quantum mechanics can not be applied to whole Universe,
 because one has the single Universe and is able to
perform one measurement. What is the sense of the scale factor
dispersion in this case?
 Our suggestion is  to consider some local system (Intermediate
System for Measurements, IMS) inside Universe and imply that
measurements are carried out under it.
 Denoting the IMS degrees of freedom as
 $\bm \xi=\{\xi_1, \xi_2 \dots\}$, one can write
 Lagrangian of the flat, uniform, and isotropic Universe including IMS:

\begin{equation}
{\mathcal L}(t)=\Omega \biggl\{ -\frac{3 M_p^2}{8\pi }a\,
\dot{a}^2  +\frac{1}{2}a^3\dot{\phi}^2- a^3V(\phi)\biggr\}+L(\bm
\xi,\dot {\bm \xi},a),
\end{equation}
where $M_p=1/\sqrt{G}$ is the Plank mass, and $\Omega$ is the
``volume'' over which the integration in the action (\ref{deistv})
is doing. IMS is local, i.e. it fills some restricted region of
the three-dimensional space which is much smaller than $\Omega
a^3$ at the moment of time considered.

 For simplicity, let us consider $\Omega$ to be infinite. As a
result, the equation of motion for Universe (including $\ddot a$)
becomes independent on the IMS variables $\bm \xi$, because the
influence of the finite system to the infinite one is negligible
(let us remind that we consider Universe with the ``frozen'' local
degrees of freedom, implying that some change of Universe occurs
in the hole space simultaneously). From the other hand, equations
for IMS contain $a(t)$ as the time-dependent parameter. After
quantization this parameter becomes the operator. For convenience
in the IMS description, we can turn to the Schr\"{o}dinger picture
and at the same time to consider the scale factor in the
quasi-Heisenberg picture as before. Thus the IMS Hamiltonian
$H(p_{\bm \xi },\bm \xi,\hat a(t))$ contains the operator $\hat
a(t)$ as a parameter. State of the system is described by the
density matrix
\begin{equation}
\rho(\bm \xi,\bm \xi',t)=<_a|\Psi(\xi',t,[\hat a])\Psi(\xi, t,
[\hat a])|_a>,
\end{equation}
where averaging over the Universe state is assumed.
 The wave
function $\Psi(\xi',t,[\hat a])$ is
 the solution of the Schr\"{o}dinger equation and
depends functionally on the operator of the Universe scale factor.

If an observable does not contain the scale factor explicitly, one
can find its mean value from the density matrix. If an observable
contains the scale factor, one has to use the wave function and to
average over the Universe state at the last step. We can consider
a number of IMSs: our consideration remains valid and the limited
number of IMSs does not influence "large" Universe. In our model
IMSs can be placed arbitrary close to each other and measurements
can be proceeded during the infinitesimal time. This does not
change the measurement results.

Let us consider the red shift from the distant sources in the
fluctuating Universe. At first, let us note that the fluctuating
scale factors does not change the spectral characteristics of the
source itself. For example, the hydrogen atom hamiltonian is
 \begin{equation}
H=\frac{\bm p_\xi^2}{2 a^2\,m} +\frac{e^2}{a\,|\bm \xi|}.
\label{hh1}
 \end{equation}
where $\xi$ is the IMS spatial coordinate.
 Here it is implied that $\xi^i\gamma_{ij}\xi^j=a^2 \bm \xi^2$ and $p_i\gamma^{ij}p_j=\frac{1}{a^2}\bm
 p^2$, where $\gamma_{ij}$ is the metric tensor of the three-space.
Let us evaluate energies  of a ``pure'' state, which is the
eigenstate of Hamiltonian (\ref{hh1}). The ``pure'' state depends
on the operator of the Universe scale factor, which is implied to
be time-independent (adiabatic approximation). To find energies we
have to rescale the mass and the charge so that $m\rightarrow a^2
m$ and $e^2\rightarrow e^2/a$, respectively. Then using the
ordinary formula
\begin{equation}
E=-\frac{m\,e^4}{2 n^2}, ~~~~~~~~n=1,2,...
\end{equation}
one can find that the energies do not change after rescaling.

When one observes the spectrum of an atom from a distant point,
the red shifted frequencies are visible in agreement with the
formula
\begin{equation}
\omega=\omega_0(1+\frac{\dot a}{a}l ),
\end{equation}
which is valid for not too large distances $l$. We can consider
the ray of light as a number of IMS situated along the ray
trajectory. Each of these IMS shows different values of the scale
factor. It is equivalent to the ray propagation in the randomly
nonuniform media. Since the scale factor is a fluctuating
quantity, the additional level width
\begin{equation}
\Gamma_a=\omega_0\, l \sqrt{\left<\left(\frac{\dot
a}{a}\right)^2\right>-\left<\frac{\dot a}{a}\right>^2}
\end{equation}
appears besides the broadening of a line due to atom collisions.

Above we considered uniform Universe. In the general case of
Universe with the local degrees of freedom, one cannot  neglect a
backreaction of the measuring system on Universe. The measurement
process spoils suggested uniformity. Under the measurements
carried out under IMS, the projection of the Universe state occurs
(Fig. \ref{stoks}, \textit{a}). The projection is local, i.e. the
Universe state is ``spoiled'' only at the local region occupied by
IMS. One does not need to consider the multiple Universes to build
quantum mechanics: every measurements spoil Universe only in a
local region, and far from this point Universe remains almost
unchanged and ready for a next measurement.

Let us discuss one more sort of measurements. There exists an
opinion \cite{kiefer}, that the matter degrees of freedom in
Universe serve as an environment leading to the decoherence of
metric. In other words, we can assume that at some stage of the
Universe evolution, a ``self measurement'' occurs due to a matter
filling Universe. As a result, the scale factor is projected to
the different values in the different spatial regions of Universe
and the nonuniform structure of Universe appears (Fig.
\ref{stoks}, \textit{b}). Quantum dispersion of the scale factor
turns to dispersion of the scale factor in the different spatial
regions (i.e. the classical dispersion results from the set of
measurements). Again, this results in an additional broadening of
spectral lines.
 If a typical size of the scale factor nonuniformity is
greater than the size of light source, it is not possible to see
an additional line broadening. In this case, another effect like
an observation of the Hubble constant dispersion in the different
directions has to appear. In the case, when the size of the
nonuniformity is greater than the observed range of Universe, we
are not able to detect the spatial scale factor dispersion by the
direct experiments.

\section{Conclusion}

Universe quantum evolution originated from the some fluctuation of
the scalar field (wave packet) has been considered. No initial
conditions for inflation are needed because all information is
contained in the quantum state.

Quantization procedure for the equation of motion  has been
introduced resulting in the quasi-Heisenberg operators, which are
Hermite when the Universe wave function normalizes in the
Klein-Gordon style.

For the quadratic inflationary potential, the numerical
calculations have demonstrated that dispersion of logarithm of the
scale factor grows during inflation and approaches some constant
at the inflation end. Interpretation of this fact can be that the
Universe scale factor is projected to the differen values within
the different spatial regions by the "self-measurement" process
(associated with the decoherence) after inflation. This results in
a highly nonuniform Universe. However,  because there is no
significant dispersion of the Hubble constant measured in a
different directions, these spatial regions must have super-Hubble
size.

In the model with massless scalar field but with the cosmological
constant, which decreases with the scale factor growth, we obtain
the decreasing dispersion of logarithm of the Universe scale
factor. This causes the negligible dispersion after the inflation
end without any need of the decoherence.

\section*{Acknowledgments}

The authors are grateful to Vladimir Baryshevsky for clarifying
discussions on measurement problem and to Konstantin Batrakov,
Eduardo Guendelman, Alexander Kaganovich, Victor Tikhomirov for
the interest to this work.

\end{document}